\magnification=1200
\null\bigskip\bigskip
\centerline{HOW SHOULD ONE DEFINE ENTROPY PRODUCTION}
\centerline{FOR NONEQUILIBRIUM QUANTUM SPIN SYSTEMS?}
\bigskip\bigskip
\centerline{by David Ruelle\footnote{*}{IHES.  91440 Bures sur Yvette,
France. $<$ruelle@ihes.fr$>$}.}
\bigskip\bigskip\noindent
	{\leftskip=2cm\rightskip=2cm\sl Abstract.  This paper discusses entropy production in nonequilibrium steady states for infinite quantum spin systems.  Rigorous results have been obtained recently in this area, but a physical discussion shows that some questions of principle remain to be clarified.\par}
\vfill\eject
\bigskip\bigskip\noindent
{\bf 1. Introduction.}
\medskip
        Recent papers by Ruelle [4], [5], and Jak\v si\'c and Pillet [3] have
discussed the nonequilibrium statistical mechanics of infinite quantum spin
systems, and in particular the positivity of entropy production.
Mathematically, these papers are based on the treasure of results accumulated
in the two volumes of {\it Operator algebras and quantum statistical mechanics}
by Bratteli and Robinson [2].  In particular the concepts of relative modular 
operator and relative entropy, developed by Huzihiro Araki [1], turn out to 
play an essential role (see [3]). Clearly, the current surge of activity in
nonequilibrium statistical mechanics is going to make more demands on operator
algebras, and on the basic structural facts discovered about these algebras by
Tomita, Takesaki, and Araki.  The present paper is however less concerned with
mathematics than with the physical question of how to define entropy
production.  We shall try to find out what is likely to be true or not true in
this area, and therefore what are the theorems that one should attempt at
proving. 
\medskip\noindent
{\bf 2. A formula for the entropy production in a finite system.}
\medskip
        In this section we consider a {\it finite} system described by a
density matrix $\psi$ on a finite-dimensional Hilbert space ${\cal H}$.  The
(von Neumann) entropy associated with $\psi$ is 
$$      S(\psi)=-{\rm Tr}\psi\log\psi      $$
In the presence of a time evolution defined by unitary operators  $U(t)$ on 
${\cal H}$ we may define 
$$      \psi(t)=U(t)\psi U(-t)      $$
It is clear and well known that the entropy $S(\psi(t))$ is independent of $t$:
there is no entropy production in this setup.  To understand entropy
production we have to think of a large system (the universe) of which we
observe a small part.  By virtue of the time evolution, there are correlations
between the state of the small system, and parts of the large system that are
more and more remote.  After a while these correlations are forgotten, or
equivalently entropy is created.  This is basically the way entropy production
was understood by Boltzmann.  
\medskip
        We shall follow this way of thinking, and consider that our system is 
composed of several subsystems labelled by an index $a=0,1,\ldots$
Corresponingly ${\cal H}=\otimes_{a\ge0}{\cal H}_a$, we assume that the
$U(t)$ form a one-paramameter group of unitary transformations of ${\cal H}$,
with $U(t)=e^{-iHt}$.
\medskip
        We can now define a density matrix $\psi_a(t)$ on ${\cal H}_a$ as a 
partial trace:
$$      \psi_a(t)={\rm Tr}_{{\cal H}_{\backslash a}}\psi(t)      $$
where ${\cal H}_{\backslash a}=\otimes_{b\ne a}{\cal H}_b$.  The
entropies 
$$      S_a(t)=-{\rm Tr}_{{\cal H}_a}\psi_a(t)\log\psi_a(t)      $$
may depend on $t$, and we define the (rate of) entropy production
$e=e(t)$ as
$$      e={d\over dt}\sum_{a\ge0}S_a(t)
        ={d\over dt}(\sum_{a\ge0}S_a(t)-S(\psi(t)))      $$
This is the rate of change of entropy associated with the decomposition
of the system described by $\psi(t)$ into the subsystems described by
$\psi_a(t)$, $a\ge0$.
\medskip
        Note that by the subbaditivity of the entropy
$$      \sum_{a\ge0}S_a(t)-S(\psi(t))\ge0      $$
This positive quantity may be viewed as the information lost about the state
$\psi(t)$ of the system when we cut it into the subsystems labelled by
$a=0,1,\ldots$  In the large system limit where correlations move away and
disappear at infinity, we expect $\sum_{a\ge0}S_a(t)-S(\psi(t))$ to be an
increasing function of $t$, so that $e\ge0$ (at least in the average).  But 
for the moment we consider finite systems, {\it i.e.}, we keep ${\cal H}$ 
finite dimensional, and we look more carefully at the expression for the 
entropy production.
\medskip
        Let us write 
$$      H=\sum_{a\ge0}H_a+h      $$
where 
$$      H_a={\bf 1}_{\backslash a}\otimes\hat H_a      $$
and ${\bf 1}_{\backslash a}$ is the unit operator on ${\cal
H}_{\backslash a}$.  (Note that that the choice of $h$, $\hat H_a$ is
not unique, one may in particular take $h=H$ and all $H_a=0$).  Then, 
to first order in $dt$, 
$$      \psi(t+dt)=e^{-iHdt}\psi(t)e^{iHdt}
        =\psi(t)-i\,dt[H,\psi(t)]
        =\psi(t)-i\,dt[\sum_{a\ge0}H_a+h,\psi(t)]      $$
hence
$$      \psi_a(t+dt)=\psi_a(t)-i\,dt[\hat H_a,\psi_a(t)]
        -i\,dt{\rm Tr}_{{\cal H}_{\backslash a}}[h,\psi(t)]      $$
Therefore, assuming that $\psi(t)$ is invertible so that the $\log$ is well
defined, 
$$  S_a(t+dt)=-{\rm Tr}_{{\cal H}_a}\psi_a(t+dt)\log\psi_a(t+dt)  $$
$$      =\sigma_1+\sigma_2      $$
where 
$$      \sigma_1=-{\rm Tr}_{{\cal H}_a}\psi_a(t)
        \log\big(\psi_a(t)-i\,dt[\hat H_a,\psi_a(t)]
        -i\,dt{\rm Tr}_{{\cal H}_{\backslash a}}[h,\psi(t)]\big)      $$
$$      =-{\rm Tr}_{{\cal H}_a}\psi_a(t)\log\psi_a(t)
        +i\,dt{\rm Tr}_{{\cal H}_a}[\hat H_a,\psi_a(t)]
+i\,dt{\rm Tr}_{{\cal H}_a}{\rm Tr}_{{\cal H}_{\backslash
a}}[h,\psi(t)]$$
$$      =-{\rm Tr}_{{\cal H}_a}\psi_a(t)\log\psi_a(t)      $$
$$      \sigma_2=i\,dt{\rm Tr}_{{\cal H}_a}([\hat H_a,\psi_a(t)]
+{\rm Tr}_{{\cal H}_{\backslash a}}[h,\psi(t)])\log\psi_a(t)      $$
$$      =i\,dt{\rm Tr}_{{\cal H}_a}({\rm Tr}_{{\cal H}_{\backslash
a}}[h,\psi(t)])\log\psi_a(t)=i\,dt{\rm Tr}_{\cal
H}([h,\psi(t)]({\bf1}_{\backslash a}\otimes\log\psi_a(t)))      $$
$$      =-i\,dt{\rm Tr}_{\cal
H}(\psi(t)[h,{\bf1}_{\backslash a}\otimes\log\psi_a(t)])      $$
Therefore
$$      S_a(t+dt)-S_a(t)=-i\,dt{\rm Tr}_{\cal
H}(\psi(t)[h,{\bf1}_{\backslash a}\otimes\log\psi_a(t)])\eqno{(1)}      $$ 
$$      e=-i\sum_{a\ge0}{\rm Tr}_{\cal
H}(\psi(t)[h,{\bf1}_{\backslash a}\otimes\log\psi_a(t)])      $$
and finally
$$      e=-i{\rm Tr}_{\cal
H}(\psi(t)[h,\log\otimes_{a\ge0}\psi_a(t)])\eqno{(2)}      $$
        Note that we may in (1) and (2) replace $h$ by the total
Hamiltonian $H$ (take all $\hat H_a=0$):
$$      e=-i{\rm Tr}_{\cal
H}(\psi(t)[H,\log\otimes_{a\ge0}\psi_a(t)])      $$
\noindent
{\bf 3. The large system limit.}
\medskip
        We shall be interested in the limit of a large system.  More
precisely, the subsystem $R_0=\Sigma$ corresponding to $a=0$ will remain
small, but it will interact with {\it reservoirs} $R_1,R_2,\ldots$, which
will become large (there is no direct interaction between the reservoirs).  
We shall be interested in the case where there are at
least two large reservoirs (in the case of only one large reservoir $R_1$, we
expect that the small system $\Sigma$ will get in equilibrium with $R_1$ if we
wait long enough -- this is the situation of {\it approach to equilibrium}).
We think of the reservoirs $R_1,R_2,\ldots$ as having different inverse
temperatures $\beta_1,\beta_2,\ldots$  Of course, putting the reservoirs in
contact with the small system $\Sigma$ will produce a flow of heat, so that
the temperature in the reservoirs will not remain uniform, in particular the
entropy production $e$ defined by the large system limit of (2) might depend on
where the boundary between the small system and the reservoirs is put.  [It is
also possible that it does not since, physically, entropy production depends
on information disappearing at infinity on different sides of some separating
surfaces, and the exact position of these surfaces may not be important].  In
any case we are interested in a double limit where first the reservoirs are
allowed to become infinite and then, perhaps, the boundaries between the small
system and the reservoirs are allowed to move to infinity.  This double limit 
is more or less imposed by physics, but seems hard to analyze mathematically.
Note for example that we expect the entropy 
$-{\rm Tr}(\psi\log\otimes_{a\ge0}\psi_a)+constant$ to diverge in a large 
system limit where it becomes time independent while its time derivative tends
to a nonzero constant $e$.  
\medskip
        We shall try to argue that in the double limit discussed above, (2)
becomes the standard thermodynamic relation between the heat fluxes and the
temperatures of the reservoirs, but we shall not be able to give a {\it proof}
of this fact.  Basically, our difficulty is to make sense of the limit of
$\log\otimes_{a\ge0}\psi_a$ or $[h,\log\otimes_{a\ge0}\psi_a]$.
\medskip\noindent
{\bf 4. Infinite systems.}
\medskip
        In order to be able to discuss a small system $\Sigma$ coupled
with actually infinite reservoirs $R_a$ with $a>0$, we shall now introduce 
more structure into the problem.  Let $L$ be countably infinite, and 
${\cal H}_x$ be a finite dimensional Hilbert space for each $x\in L$.  
We let $L$ be the disjoint union $L=\cup_{a\ge0}R_a$, where $R_0=\Sigma$
is finite and the $R_a$ with $a>0$ are infinite.  Choosing $\Lambda$ 
finite such that $\Sigma\subset\Lambda\subset L$, we may define 
${\cal H}_a={\cal H}_{\Lambda a}=\otimes_{x\in\Lambda\cap R_a}{\cal
H}_x$, ${\cal H}={\cal H}_\Lambda=\otimes_{x\in\Lambda}{\cal H}_x$, and 
study the {\it finite system} defined by a {\it density matrix} 
$\psi(t)=\psi_\Lambda(t)$ on ${\cal H}_\Lambda$ and a (self-adjoint) 
{\it Hamiltonian} $H=H_\Lambda$ on ${\cal H}_\Lambda$. 
\medskip
        For finite $X\subset L$ let ${\cal A}_X$ be the C$^*$-algebra 
of operators on ${\cal H}_X=\otimes_{x\in X}{\cal H}_x$.  If $Y\subset
X$ we may identify ${\cal A}_Y$ with a subalgebra of ${\cal A}_X$ by 
$B\mapsto B\otimes{\bf 1}_{X\backslash Y}$, and define the quasilocal 
C$^*$-algebra ${\cal A}$ corresponding to $L$ as the norm closure of 
$\cup_X{\cal A}_X$.  We can then introduce a Hamiltonian for the
infinite system $L$ as the formal expression 
$$      H_L^\Phi=\sum_{X\subset L}\Phi(X)      $$
where the sum is over finite subsets $X$ of $L$, and $\Phi(X)$ is 
self-adjoint $\in{\cal A}_X$.  The finite system Hamiltonian is then 
defined by\footnote{*}{This may be modified (for instance by boundary terms) 
provided formally $H_\Lambda\to H_L$ when $\Lambda$ tends to $L$.}
$$      H=H_\Lambda^\Phi=\sum_{X\subset\Lambda}\Phi(X)      $$
\indent
        The infinite system limit consists now in letting $\Lambda$ tend to
infinity in a suitable way, which we shall not discuss (but $\Lambda$ should
eventually contain any given finite set).  We may assume that the density 
matrices $\psi_\Lambda(t)$ tend to a time independent state $\rho$ on 
${\cal A}$ when $\Lambda\to L$ in the sense that
$$      Tr_{{\cal H}_\Lambda}\psi_\Lambda(t)A\to\rho(A)\qquad\hbox{if}\qquad
A\in{\cal A}_X      $$
for finite $X\subset L$.  We want to take for $\rho$ not just a time invariant
state, but one which qualifies as nonequilibrium steady state (so that in
particular, if the entropy production can be defined, it is not negative).  We
shall discuss nonequilibrium steady states below.
\medskip
        Of the quantities occuring in (1) and (2) we see that we can now
replace ${\rm Tr}_{\cal H}(\psi(t)\cdots)$ by $\rho(\cdots)$.  For {\it finite
range} interactions, $h$ is a well defined element of ${\cal A}$ and
independent of $\Lambda$ for sufficiently large $\Lambda$.  The operator $H$
is, in the limit of infinite $\Lambda$ given formally by $H_L^\Phi$ as defined
above.  
\medskip
        It is however not clear what to do with the limit of
$\log\otimes_{a\ge0}\psi_a(t)$.  One idea would be to  assume that 
$$      \log\psi_\Lambda(t)+c_\Lambda{\bf 1}_\Lambda\quad
        \to\quad-\sum_{X\subset L}\Psi(X)      $$
where the $c_\Lambda$ are constants and the right hand side is (up to
sign) a formal sum of self-adjoint elements $\Psi(X)\in{\cal A}_X$ for
finite $X\subset L$.  But such an Ansatz conflicts with the notion that
$\log\psi_a(t)$ for a reservoir has long distance correlations, {\it i.e.},
very large or infinite sets $X$ should be important in the the formula
displayed above.  In conclusion, we believe, for physical reasons that the
infinite system limit of the entropy production makes sense, but we cannot
prove this fact.
\medskip\noindent
{\bf 5. The thermodynamic formula for the entropy production.}
\medskip
        At this point we have come to an expression of $e$ as limit when
$\Lambda$ tend to infinity (or $\Lambda\to L$) of
$$      -i{\rm Tr}(\rho_\Lambda[H_\Lambda^\Phi,\log\otimes_{a\ge0}\psi_a])
=i{\rm Tr}([H_\Lambda^\Phi,\rho_\Lambda]\log\otimes_{a\ge0}\psi_a)      $$
Since $\rho$ is invariant under the time evolution defined by $H_L^\Phi$, we
have $\rho([H_\Lambda^\Phi,A]=0$ if $A$ belongs to a local algebra and 
$\Lambda$ is sufficiently large.  Therefore, 
${\rm Tr}([H_\Lambda^\Phi,\rho_\Lambda]A)$ vanishes if $A$ is localized well inside 
$\Lambda$, and is nonzero only for $A$ localized near the boundary of 
$\Lambda$.  In other words, in computing $e$ we may ignore local terms from 
$\log\otimes_{a\ge0}\psi_a$ and pay attention only to contributions from far 
away, at the boundary of $\Lambda$.  An obvious guess is then to replace 
$\psi_a$ by the equilibrium state at temperature $\beta_a$ in $R_a$, obtaining
now $e$ as limit when $\Lambda\to L$ of 
$$      i{\rm Tr}(\rho_\Lambda[H_\Lambda^\Phi,\sum_a\beta_a 
        H_{\Lambda\cap R_a}^\Phi])
        =i\rho(\sum_a\beta_a[H_\Lambda^\Phi,H_{\Lambda\cap R_a}^\Phi])      $$
Note that $\lim_{\Lambda\to L}i[H_\Lambda^\Phi,H_{\Lambda\cap R_a}^\Phi]$ is a
well defined operator localized near the surface of the small system $\Sigma$,
it represents the rate of transfer of energy to the reservoir $R_a$ and
therefore for large enough $\Lambda$
$$      e=\rho(\sum_a\beta_a i[H_\Lambda,H_{\Lambda\cap R_a}])\eqno{(3)}  $$
In fact (3) is the usual thermodynamic expression of the
entropy production in terms of heat fluxes.  Note that we may ignore the term
with $a=0$ since the fluxes to the small system add up to 0 in a stationary
state.  
\medskip
        We shall from now on proceed with the formula (3) for the entropy
production, but remember that its relation with (2) has not been
satisfactorily established.
\medskip\noindent
{\bf 6. Nonequilibrium steady states {\rm (NESS)}.}
\medskip
        The definition of nonequilibrium steady states (NESS) should choose a
direction of time, {\it i.e.}, distinguish between the past and the future.
Otherwise one cannot hope to prove that the entropy production $e$ has a
definite sign.  One must also impose the asymptotic temperatures
$\beta_1^{-1},\beta_2^{-1},\ldots$ in the reservoirs $R_1,R_2,\ldots$  We
assume that the interaction $\Phi$ determines a one-parameter group
$(\alpha^t)$ of automorphisms of ${\cal A}$, defining the time evolution of
our system (see [2]).  Let $\sigma_1,\sigma_2,\ldots$ be equilibrium states at
inverse temperature $\beta_1,\beta_2,\ldots$ for the reservoirs
$R_1,R_2,\ldots$ and $\sigma_0$ any state for the small system $\Sigma=R_0$.
Assuming the existence for each $A\in{\cal A}$ of a limit
$$ \lim_{t\to\infty}(\otimes_{a\ge0}\sigma_a)(\alpha^tA)=\rho(A)\eqno{(4)} $$
defines a state $\rho$ which one can certainly call a NESS.  One can prove the
existence of the limits (4) under strong conditions of asymptotic abelianness
in time of the evolution $(\alpha^t)$ and also of the ``uncoupled'' evolutions
$(\breve\alpha_a^t)$.  This was the point of view adopted in [4].  It has the
advantage of leading to strong results like linear response formulae, but the
disadvantage that the assumed asymptotic abelianness can practically never be
verified.  Progress in understanding nonequilibrium quantum spin systems will
probably depend on a better understanding of the asymptotic abelianness
conditions in question.  
\medskip
        It is however possible to obtain some results without unverifiable
assumptions by considering limit points for $T\to\infty$ of 
$$ {1\over T}\int_0^Tdt\,(\alpha^t)^*(\otimes_{a\ge0}\sigma_a)\eqno{(5)} $$
in the weal dual of ${\cal A}$.  Such limit points $\rho$ always exist (by
$w^*$-compactness of the set of states) and they are invariant under time
evolution.  The limit points $\rho$ are good candidates to represent
nonequilibrium steady states.  In fact, it has been proved in [5], and more
generally in [3] that the entropy production $e$ defined by (3) is $\ge0$ for
such states.  [The reason is basically that the commutator in (3) is a
derivative, which combines with the integral in (5) to produce a manageable
expression].  
\medskip
        If one assumes that the $\sigma_a$ ($a>0$) are extremal KMS states and
that $(\alpha^t)$ is asymptotically abelian one can show (see [5]) that the
definition of the NESS as limit points of (5) does not depend on where the
boundaries between the small system $\Sigma$ and the reservoirs $R_a$ ($a>0$)
are placed.  This is of course quite desirable.  Asymptotic abelianness of
$(\alpha^t)$ also ensures that the NESS have a unique ergodic decomposition.
So, even with the definition of NESS based on (5), questions of asymptotic
abelianness seem to appear unavoidably.  This is natural because if our system
is composed of subsystems that do not interact (say the small system does not
interact with the reservoirs, we violate asymptotic abelianness, and have an
uninteresting theory.
\medskip        
        We have said nothing of the geometry of the reservoirs $R_a$ ($a>0$),
but consideration of the macroscopic limit shows that (if they are pieces of
regular lattices ${\bf Z}^d$) their dimension $d$ must be $\ge3$.  Indeed in
the macroscopic limit, a NESS corresponds to a temperature field $T$
satisfying $\triangle T=0$ (say), and tending to limits $\beta_a^{-1}$ in the
various reservoirs.  In view of properties of harmonic functions this cannot
happen for $d<3$.  What will happen for $d<3$ is that the temperature will
tend, as time goes to infinity, to a constant in any bounded region, the
temperature gradient and heat flux will tend to zero, and the NESS will reduce
to a thermodynamic equilibrium state with $e=0$.
\medskip
	In conclusion we hope to have shown in this note that, on our way to 
understanding quantum nonequilibrium statistical mechanics, there remains not 
only problems of mathematics to solve but also questions of physics to clarify.
\vfill\eject
\noindent
{\bf References.}

[1] H. Araki  ``Relative entropy of states of von Neumann algebras''
Publ. R.I.M.S., Kyoto Univ. {\bf 11},809-833(1976).

[2] O. Brattelli and D.W. Robinson  {\it Operator algebras and quantum
statistical mechanics}, Springer-Verlag, Berlin, 2-nd ed. I(1987),II(1996).

[3] V. Jak\v si\'c and C.-A. Pillet  ``On entropy production in quantum
statistical mechanics''  Preprint.

[4] D. Ruelle  ``Natural nonequilibrium states in quantum statistical
mechanics'' J. Statist. Phys. {\bf 98},57-75(2000).

[5] D. Ruelle  ``Entropy production in quantum spin systems''  Preprint.
\end